\documentclass[aps,showpacs,prl,twocolumn]{revtex4-2}  

\usepackage[retainorgcmds]{IEEEtrantools} 

\usepackage{amsmath}
\usepackage{mathrsfs}
\usepackage{amssymb}
\usepackage{graphicx,epstopdf,epsfig}
\usepackage{bm}
\usepackage{color}
\usepackage{times}
\usepackage{hyperref}
\usepackage{physics}

\definecolor{clr}{rgb}{0,0.6,0.6}

\newcommand{\mr}[1]{{\mathrm{#1}}} 		

\renewcommand{\Re}{\mr{Re}}			

\newcommand{\etal}{\,\textit{et al.~}}

\newcommand{\SM}{\,\cite{SM}}

\usepackage{tikz}
\usetikzlibrary{quantikz}
\usepackage{nicefrac}

\begin{document}

\title{Demonstration of a Quantum Gate using Electromagnetically Induced Transparency}

\author{K. McDonnell}
\author{L.F. Keary}
\author{J.D. Pritchard}
\email{jonathan.pritchard@strath.ac.uk}
\affiliation{EQOP, Department of Physics, University of Strathclyde,  SUPA, Glasgow G4 0NG, UK}

\date{\today}
\begin{abstract} 
We demonstrate a native $\mathrm{CNOT}$ gate between two individually addressed neutral atoms based on electromagnetically induced transparency (EIT). This protocol utilizes the strong long-range interactions of Rydberg states to enable conditional state transfer on the target qubit when operated in the blockade regime. An advantage of this scheme is it enables implementation of multi-qubit CNOT$^k$ gates using a pulse sequence independent of qubit number, {\color{black}{providing a simple gate for efficient implementation of digital quantum algorithms and stabiliser measurements for quantum error correction}}. We achieve a loss corrected gate fidelity of $\mathcal{F}_\mathrm{CNOT}^\mathrm{cor} = 0.82(6)$, and prepare an entangled Bell state with $\mathcal{F}_\mathrm{Bell}^\mathrm{cor} = 0.66(5)$, limited at present by laser power. We present a number of technical improvements to advance this to a level required for fault-tolerant scaling.
\end{abstract}

\maketitle

Neutral atoms are a promising candidate for scalable quantum computing, pairing long coherence times with strong long-range interactions of highly excited Rydberg states \cite{saffman10,adams19,henriet20,morgado21,cong22}. Major advantages over other technologies are the ease with which the system can be scaled to create deterministically loaded, defect-free arrays of single atoms in 1D \cite{endres16}, 2D {\color{black}{\cite{kim16,barredo16}}} and 3D {\color{black}{\cite{lee16,barredo18}}} or through use of atomic ensembles \cite{ebert15}. Additionally the exquisite control over the atom-atom interactions offered through choice of Rydberg state and tuning using external static \cite{ravets15}, microwave \cite{tanasittikosol11,maxwell13,sevincli14} or optical \cite{leseleuc17} electric fields to engineer highly anisotropic interactions with variable length scale. The strong long-range interactions give rise to a blockade mechanism whereby within a volume of radius $R\lesssim10~\mathrm{\mu}$m only a single Rydberg excitation can be created \cite{lukin01}. Rydberg blockade can be exploited to create deterministic entanglement \cite{wilk10,picken18,levine18,madjarov20} or realize high-fidelity two-qubit gate operations \cite{isenhower10,maller15,zeng17,graham19,levine19,fu22}. These gates have enabled recent demonstration of quantum algorithms \cite{graham22}, with the ability to engineer non-local qubit connectivity through use of mobile tweezers to dynamically rearrange atoms \cite{bluvstein22}.

The strong interactions can be further extended to perform native multi-qubit gates \cite{brion07a,isenhower11,beterov18,shi18,li21,su18,young21,wu10,wu10a,su18a,khazali20,rasmussen20,muller09} providing a route to efficient implementation of quantum circuits \cite{molmer11,petrosyan16}. These gates can be realized using sequential excitation pulses applied to each qubit \cite{isenhower11,beterov18,shi18,li21}, or through simultaneous addressing \cite{wu10,khazali20} as recently demonstrated for a three-qubit Toffoli gate \cite{levine19}. For both approaches, pulse-shaping and quantum optimal control techniques have been utilized to obtain high-fidelity protocols \cite{levine19,pelegri21,jandura22}. However, due to the $\sqrt{N}$ scaling of the collective Rabi frequency for Rydberg excitation in the blockade regime, these protocols require re-optimisation as the number of qubits changes. An alternative approach based on electromagnetically induced transparency (EIT) was originally proposed by M\"uller \emph{et al.} \cite{muller09}. This scheme provides a scalable approach to performing multi-qubit gates with a single control and $k$ target qubits (CNOT$^k$) without the need to compensate for the collectively enhanced Rabi frequency, enabling implementation using a pulse-sequence that is independent of $k$.

In this paper we present {\color{black}{the first}} demonstration of this EIT gate protocol for two-qubits, verifying the ability to perform a native $\mathrm{CNOT}$ gate without requiring additional single qubit rotations, yielding a loss corrected gate fidelity of $\mathcal{F}_\mathrm{CNOT}^\mathrm{cor} = 0.82(6)$. We utilize this gate sequence to prepare an entangled Bell state with a corrected fidelity of $\mathcal{F}_\mathrm{Bell}^\mathrm{cor} = 0.66(5)$. Whilst our current demonstration is limited by laser power, we propose a number of technical improvements to reach sufficiently high fidelities to facilitate the creation of entangled states for performing measurements beyond the standard quantum limit \cite{maccormick16}, and achieving fault-tolerant computing using surface codes  for topological error correction \cite{auger17}. 

\begin{figure}
    \centering
    \includegraphics{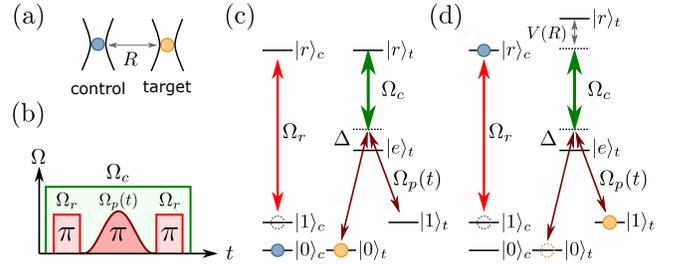}
    \caption{EIT gate protocol. (a) A single control and target atom trapped in two optical dipole traps separated by distance $R$. (b) $\mathrm{CNOT}$ pulse sequence. (c) If the control atom is in $\ket{0}_c$ during the smooth pulse the target qubit adiabatically follows the EIT dark state $\ket{0}_t\rightarrow\ket{0}_t$ leaving its state unchanged. (d) If the control atom is initially in $\ket{1}_c$ then strong dipole-dipole interactions $V(R)$ detune the target qubit Rydberg state, breaking the EIT resonance and enabling resonant transfer $\ket{0}_t\rightarrow\ket{1}_t$.} 
     \label{fig:EIT_Protocol}
\end{figure}

The $\mathrm{CNOT}$ gate protocol proposed by M\"uller $\mathit{et~al.}$~{\color{black}{\cite{muller09}}} is illustrated in Fig.~\ref{fig:EIT_Protocol}, with control and target atoms with states $\ket{i}_{c,t}$ respectively where $i=0,1$ correspond to computational basis states and $i=e,r$ are the intermediate excited and Rydberg states respectively. We consider two atoms that are optically trapped at a separation $R$ (Fig.~\ref{fig:EIT_Protocol}(a)), and individually addressed. The control qubit is coupled from $\ket{1}_{c}\rightarrow\ket{r}_{c}$ by a laser with Rabi frequency $\Omega_r$ and the target qubit is addressed by a pair of ground-state Raman lasers each with Rabi-frequency $\Omega_p(t)$ driving a two-photon resonance from $\ket{1}_t\rightarrow\ket{0}_t$ with detuning $\Delta$ from the intermediate excited state $\ket{e}_t$. A strong coupling laser with detuning $-\Delta$ couples $\ket{e}_t\rightarrow\ket{r}_t$ with Rabi frequency $\Omega_c$. The temporal excitation sequence for the $\mathrm{CNOT}$ gate protocol is shown in Fig.~\ref{fig:EIT_Protocol}(b), where a $\pi$-pulse is applied to the control qubit followed by a smooth adiabatic pulse with area {\color{black}{$A\!=\!\int\!\mathrm{d}t \Omega_p(t)^2/2\Delta\!=\!\pi$}} on the target qubit, then a final $\pi$-pulse on the control qubit. 

For the case of the control atom initially in state $\ket{0}_c$ as shown in Fig.~\ref{fig:EIT_Protocol}(c), the Hamiltonian for the target qubit is given by
\begin{equation}
\begin{split}
\mathcal{H}_t &= \hbar\Omega_p(t)/2\left(\ket{1}_t\!\bra{e}+\ket{0}_t\!\bra{e}\right)\\&+\hbar\Omega_c/2\ket{e}_t\!\bra{r}-\hbar\Delta\ket{e}_t\!\bra{e} + \mathrm{h.c.},
\end{split}
\end{equation}
which for $\vert\Delta\vert\gg\Omega_p(t),\Omega_c$ allows adiabatic elimination of the intermediate $\ket{e}_t$. The resulting Hamiltonian has two EIT dark states $\ket{d_1}_t=(\ket{1}_t-\ket{0}_t)/\sqrt{2}$ and $\ket{d_2}_t=(1+x^2)^{-1/2}\left[(\ket{1}_t+\ket{0}_t)/\sqrt{2}-x\ket{r}_t\right]$ with $x=\sqrt{2}\Omega_p(t)/\Omega_c$ \cite{muller09}. 

For $\Omega_c/\Omega_{p}^{\mathrm{max}} \gtrsim 2$, and with the target qubit initially in an arbitrary state $\ket{\psi}_t = \alpha\ket{d_1} + \beta \ket{d_2}$, during the smooth Raman pulse the qubit adiabatically follows the dark state corresponding to $\ket{0}_c\ket{\psi}_t \rightarrow \ket{0}_c\ket{\psi}_t$.

If instead the control qubit is in state $\ket{1}_c$, the initial $\pi$-pulse transfers population to $\ket{r}_c$ resulting in detuning of the target Rydberg state by the dipole-dipole interaction energy $V(R)$ as shown in Fig.~\ref{fig:EIT_Protocol}(d). This modifies the target Hamiltonian to $\mathcal{H}'_t  = \mathcal{H}_t +V(R)\ket{r}_t\!\bra{r}$, which for $V(R)>\hbar\Omega_c^2/(4\Delta)$ is sufficient to break the EIT condition enabling the target qubit to undergo a Raman $\pi$-pulse. This protocol thus realizes a native $\mathrm{CNOT}$ gate corresponding to the mapping $\ket{1}_c\ket{0}_t \leftrightarrow \ket{1}_c\ket{1}_t$.

\begin{figure}
    \centering
    \includegraphics{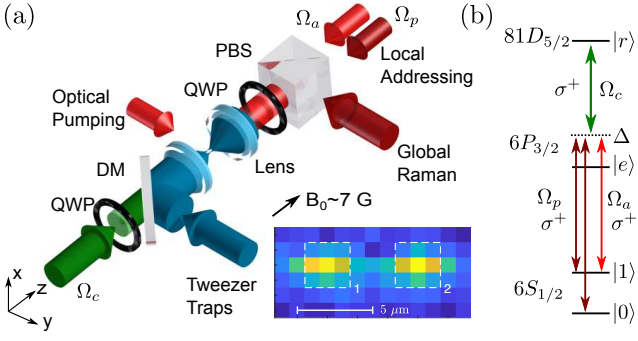}
        \caption{\small{{\color{black}{Experiment Setup. (a) Schematic}} showing single atoms trapped in microscopic tweezer traps, overlapped with the Rydberg laser on a dichroic mirror (DM) and circularly polarized using a quarter wave plate (QWP). The Raman and qubit lasers are combined on a polarizing beam splitter (PBS) and counter propagate with the Rydberg and trapping lasers. (b) Qubit level scheme with $\ket{1} = \ket{6S_{1/2}, F=4, m_F = 0}$,  $\ket{0} = \ket{6S_{1/2},F=3, m_F = 0}$  and $ \ket{r} = \ket{81D_{5/2}, m_{j}=5/2}$. The qubit (red) and Rydberg (green) lasers drive a two-photon transition from $\ket{1}\rightarrow \ket{r}$ detuned by $\Delta$ from the intermediate state $\ket{e}$. The Raman laser $\Omega_p$ (brown) drives transitions between $\ket{1} \rightarrow\ket{0}$, and is phase locked to $\Omega_a$.}}
    \label{fig:experiment}
\end{figure}

The experiment setup, previously described in reference \cite{picken18} and illustrated schematically in Fig.~\ref{fig:experiment}(a), uses a pair of individually trapped $^{133}$Cs atoms separated by $6~\mu$m that are cooled to 5~$\mu$K and detected using fluorescence collected on a sCMOS camera~\cite{Picken17}. {\color{black}{Qubits are encoded in the hyperfine clock states, with $\ket{1}=\ket{6S_{1/2},F=4, m_{F}=0} $ and $\ket{0}=\ket{6S_{1/2},F=3, m_{F}=0}$. Atoms are prepared in $\ket{1}$ using a linearly polarized optical pumping beam resonant with the transition from $\ket{6S_{1/2},F=4}\rightarrow\ket{6P_{1/2},F'=4}$}}, and we implement destructive state detection using a strong resonant blow-away beam to remove atoms in $F=4$ prior to imaging. To overcome limitations in optical pumping fidelity due to finite polarisation purity, after preparation in $\ket{1}$ we apply a resonant microwave $\pi$-pulse from $\ket{1}\rightarrow\ket{0}$ followed by a blow-away pulse to eject atoms from {\color{black}{$F=4$}} to eliminate {\color{black}{errors from}} states outside the computational basis \SM{}. {\color{black}{To suppress AC Stark shifts and Rydberg anti-trapping from the dipole trap beam, trap light is extinguished for 5~$\mu$s during which the EIT gate pulses are applied.}}

For demonstration of the EIT gate protocol we utilize the laser couplings shown in Fig.~\ref{fig:experiment}(b). Rydberg excitation to 81$D_{5/2}$ is performed using two-photon excitation via the $6P_{3/2}$ intermediate excited state. An 852~nm diode laser with Rabi frequency {\color{black}{$\Omega_a$}} couples $\ket{1}_c\rightarrow\ket{e}_c$ with detuning $\Delta$, and a frequency doubled Ti:Sapph laser at 509~nm with detuning $\Delta_c\sim-\Delta$ and Rabi frequency $\Omega_c$ couples from $\ket{e}_c\rightarrow\ket{r}_c$. These lasers are locked to an ultra-low expansion (ULE) cavity to obtain sub-kHz linewidths~\cite{legaie18} using a detuning of $\Delta/2\pi = 870$~MHz from the $\ket{6S_{1/2},F=4}\rightarrow\ket{6P_{3/2},F'=5}$ transition. The qubit laser $\Omega_a$ is focused to a waist of $3~\mu$m to locally address the control qubit, whilst the coupling laser is focused down to a $1/e^2$ waist of $18~\mu$m to illuminate both control and target atoms equally. Both beams are $\sigma^+$ polarized to maximize coupling from $\ket{1}\rightarrow\ket{81D_{5/2},m_j=5/2}$ resulting in a two-photon Rabi frequency of $\Omega_r/2\pi = 1.77$~MHz.

The Raman laser driving two-photon couplings from $\ket{1}\rightarrow\ket{0}$ is derived from a second 852~nm diode laser which uses an electro-optic modulator to generate sidebands at $\pm4.6~$GHz before filtering out the carrier using a Mach–Zehnder interferometer \cite{haubrich00} to obtain co-propagating Raman beams with equal amplitude \cite{picken18}. To ensure the Raman laser meets the EIT resonance condition, the carrier is phase-locked to the qubit laser ($\omega_a$) to transfer the narrow linewidth whilst ensuring both target and control atoms have a common intermediate state detuning with the coupling laser. The Raman light is aligned onto the target qubit using a tightly focused beam waist of $3~\mu$m with Rabi frequency $\Omega_p(t)$ and $\sigma^+$ polarisation to perform local operations and EIT. A second orthogonally polarized Raman beam with a $1/e^2$ waist of $15~\mu$m is used to perform global operations on both qubits. This beam has an additional 80~MHz detuning from the intermediate level to avoid creating additional EIT resonances with states $m_j=+1/2,3/2$.

\begin{figure}[t!]
    \centering
    \includegraphics{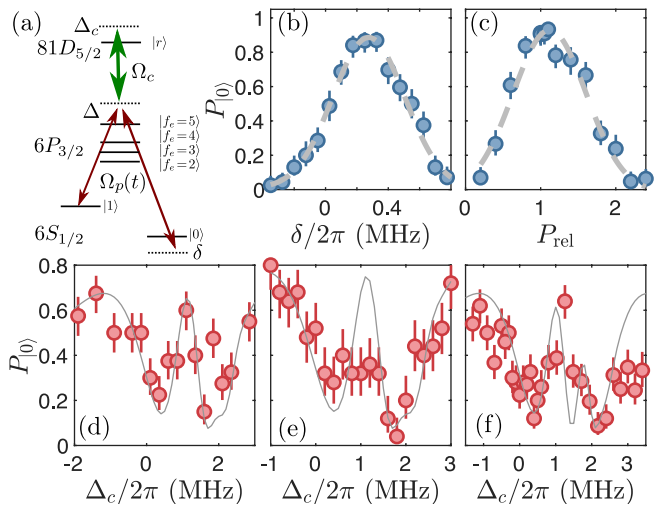}
    \caption{Pulse Optimisation. (a) Target atom qubit excitation scheme, showing all the hyperfine levels of the intermediate state $6P
    _{3/2}$. (b) Smooth-pulse optimisation with $\Omega_c=0$ to maximise transfer $\ket{1}\rightarrow\ket{0}$ as a function of two-photon detuning $\delta$ and (c) relative pulse power {\color{black}{$P_\mathrm{rel}$}} for $\tau=2~\mu$s.  (d-f) EIT optimisation vs coupling laser detuning for $\tau = 1.5,2,3~\mu$s to find EIT resonance where state transfer is suppressed due to adiabatic following of the dark state $\ket{1}\rightarrow\ket{1}$. Data are overlaid with theoretical model (grey line) \SM{}.}
    \label{fig:EIT}
\end{figure}

To implement the CNOT protocol we use an acousto-optical modulator to apply a smooth adiabatic pulse to the target qubit of the form $\Omega_p(t) = \Omega^\mathrm{max}_p(t) (1-\cos(2\pi t/\tau))/2$, resulting in a pulse area of $A = 3\tau\Omega_p^\mathrm{max}/8$ where $\tau$ is the pulse duration and $\Omega_p^\mathrm{max}$ is the peak two-photon Rabi frequency from $\ket{0}_t\rightarrow\ket{1}_t$. The description above presents a simplified picture as two-photon excitation via $6P_{3/2}$ involves not one but four intermediate hyperfine states $\ket{f_e,m_f=1}_t$ coupled to the target qubit (Fig.~\ref{fig:EIT}(a)). Only $f_e=3,4$ contribute to the Raman and EIT resonance, whilst the $f_e=2$ and $5$ states provide independent routes for Rydberg excitation from $\ket{1}$ or $\ket{0}$ and contribute significant AC Stark shifts to the Raman resonance. Whilst the AC shifts evolve dynamically during the pulse sequence, we find applying a fixed detuning of the Raman laser is sufficient to obtain high fidelity state transfer. To optimize pulse parameters the smooth adiabatic Raman pulse is applied in the absence of the coupling laser and the peak power and detuning adjusted to maximize the state transfer $\ket{1}_t\rightarrow\ket{0}_t$, with results shown in Fig.~\ref{fig:EIT}(b,c). For a $\tau=2~\mu$s pulse duration the total peak power in the Raman beam is 110~nW, corresponding to $\Omega^\mathrm{max}_p/2\pi =0.67$~MHz and an optimal Raman detuning of $\delta/2\pi = 0.28(2)$~MHz in excellent agreement with theory \SM{}. 

Following optimisation of coherent state transfer, the coupling laser is then applied and its detuning is scanned to locate the EIT resonance corresponding to the frequency at which $\ket{1}_t\rightarrow\ket{1}_t$. Data in Fig.~\ref{fig:EIT}(d-f) show EIT scans as a function of pulse duration for $\tau=1.5, 2$ and 3$~\mu$s, taken with a coupling power of 170~mW corresponding to a coupling Rabi frequency of $\Omega_c/2\pi\sim40$~MHz for $f_e=3,4$. In each case, the data are compared to numeric simulations with good qualitative agreement with the theoretical model {\color{black}{with additional features in the spectra coming from the hyperfine structure of the intermediate $6P_{3/2}$ level}} \SM{}. For $\tau=1.5~\mu$s the resulting EIT is not well defined, with leakage to $\ket{1}_t$ at all detunings due to the finite ratio $\Omega_c/\Omega_p\lesssim1$. {\color{black}{For longer pulse durations we observe suppression of state transfer on the EIT resonance. Below we use $\tau=2~\mu$s to minimize the time the control qubit is required to remain in the Rydberg state. For this duration with {\color{black}{$\Delta_c/2\pi=1.8$~MHz}} we measure $P_{\ket{0}_t}=0.04(10)$, showing minimal leakage during the adiabatic pulse evolution.}}

Using the optimized pulse parameters on the target qubit we proceed to demonstrate the $\mathrm{CNOT}$ gate operation applied to the two qubits, where for a 6~$\mu$m separation $V(R)/2\pi = 35$~MHz \cite{sibalic17}. To characterize the gate operation we prepare atoms in each of the four computational basis states using microwave pulses and measure the resulting output states. Local microwave operations are implemented using the method of Xia \etal{} \cite{xia15} by applying a calibrated AC Stark shift on the control qubit to ensure it undergoes a $4\pi$ rotation during the target qubit operation \SM{}. Using destructive blow-away it is not possible to discriminate between atom loss events and an atom in $F=4$ being removed from the trap. To overcome this issue, after the gate is applied we rotate each of the basis states into $\ket{00}$ to allow measurements conditioned on two-atom survivals, and the resulting corrected output probabilities are obtained by rescaling the raw two-atom survivals by the two-atom survival probability when no blow away beam is applied. We characterise our ability to prepare and measure computational states using this technique, resulting in raw and corrected state preparation fidelities given by $\mathcal{F}_\mathrm{prep}=1/4\mathrm{Tr}(U^T_\mathrm{meas}*U_\mathrm{ideal})=0.81(2)$ and $0.91(4)$ respectively and an average two-atom survival probability of 0.89(1) \SM{}.

\begin{figure}[t!]
    \centering
    \includegraphics{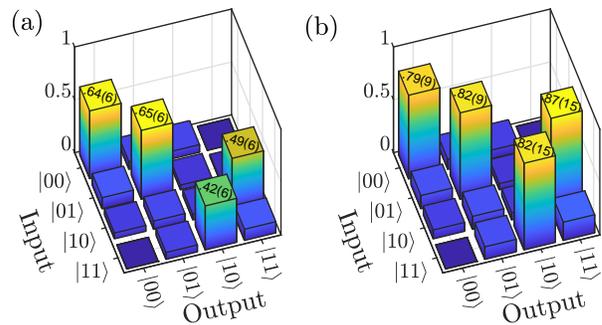}
    \caption{Gate Measurement. (a) Raw and (b) loss-corrected CNOT gate data with fidelities of $\mathcal{F}_\mathrm{CNOT}=0.55(3)$ and $\mathcal{F}_\mathrm{CNOT}^\mathrm{cor}= 0.82(6)$.}
    \label{fig:gate}
\end{figure}

The EIT gate matrix is shown in Fig.~\ref{fig:gate}, with the uncorrected measurements clearly revealing the characteristic structure of the native $\mathrm{CNOT}$ gate and verifying that for the control qubit in $\ket{0}_c$ the EIT on the target site maintains its initial state whilst for the control atom in $\ket{1}_c$ we obtain a rotation of the target states. From the raw data we see the probability of survival is reduced by $\sim25~\%$ when the control atom is excited to the Rydberg state due to additional losses of the control atom from the Rydberg state. This loss is dominated by the finite laser phase noise which reduces the probability of the control qubit returning to $\ket{1}_c$ after the two $\pi$-pulses \cite{levine19}, and is much larger than the $<5~\%$ loss predicted from radiative decay and off-resonant scattering from the coupling laser during the $\tau=2~\mu$s the control atom is in the Rydberg state \SM{}. The raw gate fidelity is $\mathcal{F}_\mathrm{CNOT}=0.55(3)$, and in Fig.~\ref{fig:gate}(b) we show that renormalizing the elements by the two-atom survival probability results in comparable values for all non-zero elements and a corrected gate fidelity of $\mathcal{F}_\mathrm{CNOT}^\mathrm{cor}= 0.82(6)$.

\begin{figure}
    \centering
    \includegraphics{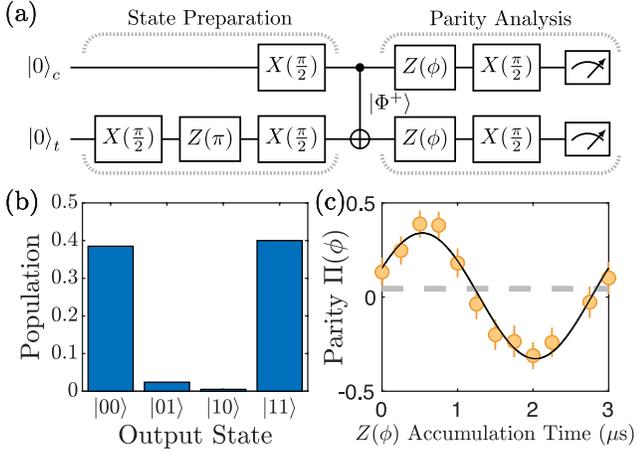}
    \caption{Bell state preparation (a) Gate sequence applied for Bell state preparation and analysis. (b) Measured Bell state populations. (c) Parity oscillation with amplitude $\vert c \vert = 0.17(1) $.}
    \label{fig:parity}
\end{figure}

To demonstrate the $\mathrm{CNOT}$ protocol is able to generate deterministic entanglement we prepare atoms in the $\ket{\Phi^+}=(\ket{00}+\ket{11})/\sqrt{2}$ Bell state using the gate sequence shown in Fig.~\ref{fig:parity}(a). For these measurements state preparation is performed using the Raman lasers rather than microwaves to ensure the phase of the input state is well defined with respect to the phase of the Raman pulse applied during the gate. We apply a local $X(\pi/2)$ to the target qubit, followed by a global $X(\pi/2)$ pulse. The delay between pulses is chosen such that the target qubit accumulates phase $Z(\pi)$ to map $\ket{00}\rightarrow (\ket{00}+i\ket{10})/\sqrt{2}$ \SM{} which is converted to $\ket{\Phi^+}$ following application of the $\mathrm{CNOT}$ gate. Bell state populations are shown in Fig.~\ref{fig:parity}(b), with direct measurement of $\rho_{00}$ ($\rho_{11}$) performed from measurement of two-atom survivals with (without) a global $X(\pi)$ pulse applied, and the remaining elements estimated using the lower bound of \cite{levine18}. 

The fidelity of the generated Bell state is equal to $\mathcal{F}_\mr{Bell} = \bra{\Phi^+}\rho\ket{\Phi^+}=(\rho_{00}+\rho_{11})/2+\vert c\vert$, where $c=\vert c\vert e^{i\phi_c}$ is the coherence between $\ket{00}$ and $\ket{11}$. The coherence is measured using parity oscillations after a global phase accumulation $Z(\phi)$ and global rotation $X(\pi/2)$, where the phase accumulation is realized by varying the delay prior to the final analysis pulse \cite{sackett00}. The resulting parity $\Pi(\phi)=\rho_{00}+\rho_{11}-\rho_{01}-\rho_{10}=2\Re(d)-2\vert c\vert \cos(2\phi+\phi_c)+\rho_{xx}$, where $d$ is the coherence between $\ket{01}$ and $\ket{10}$ and $\rho_{xx}$ is the two-atom loss probability \SM{}. Fig.~\ref{fig:parity}(c) shows the measured parity oscillation, corresponding to a Bell state coherence with amplitude $\vert c\vert = 0.17(3)$ and an average value $\langle \Pi\rangle_\phi = 0.04(2)$ in good agreement with $\rho_{xx}=0.06(2)$ measured independently in the absence of the state-selective blow away beam.

Combining the measurements of population and coherence, we find a raw fidelity of $\mathcal{F}_\mr{Bell}=0.44(5)$ which lies below the threshold for entanglement and below the theoretically predicted value of 0.78 \SM{} due to the enhanced losses discussed above. However, from our loss-corrected population measurements, and rescaling the coherence by the average two-atom survival without blow-away $P=0.67(5)$, we obtain a corrected fidelity of $\mathcal{F}^\mathrm{cor}_\mr{Bell}=0.66(7)$, demonstrating the gate protocol is capable of generating entangled quantum states.

In conclusion, we have presented the first realization of a native $\mathrm{CNOT}$ gate between two neutral atoms based on EIT. Correcting for losses we obtain a gate fidelity $\mathcal{F}_\mathrm{CNOT}^\mathrm{cor} = 0.82(6)$ and demonstrate coherent parity oscillations of a $\ket{\Phi^+}$ Bell state achieving $\mathcal{F}_\mr{Bell}^\mathrm{cor}=0.66(7)$. The major limitations in the current implementation arise from technical dephasing noise in the Rydberg excitation lasers which prevent high fidelity recovery of the control atom from the Rydberg state, and the limited power available for the coupling laser which reduces the gate speed, increasing the time the control qubit must remain in the Rydberg manifold. 

This limitation can be circumvented using two-photon excitation via 7$P_{1/2}$, which benefits from reduced scattering and AC Stark shift errors due to having fewer intermediate hyperfine levels,  a reduction in excited state linewidth $\sim\times 1/5$ and can utilise high-power laser sources at 1039~nm. We show that for modest parameters $\mathcal{F}>0.998$ are achievable for 500~ns gate times \SM{}. This, combined with techniques to suppress laser phase noise \cite{levine18}, offers a route to high-fidelity gate implementation competitive with current controlled phase gates \cite{levine19,graham19,fu22}. The primary advantage of the EIT protocol demonstrated here is the intrinsic scalability to many qubits, enabling implementation of an identical pulse sequence on $k$ target qubits to realize a $\mr{CNOT}^k$ which provides an important gate for error correction \cite{auger17}. {\color{black}{Multi-qubit gate fidelities are limited by residual target-target interactions, however this can be suppressed using a combination of geometric arrangement \cite{leseleuc19a}, careful choice of states \cite{khazali20} or different atomic isotopes \cite{zeng17} or species \cite{beterov15} to obtain a fidelity of $\mathcal{F}_k=0.998^k$ \SM{}.}

\emph{Note added}: During completion of our manuscript we became aware of related work calculating multi-qubit gate fidelities based on this EIT protocol using heteronuclear interactions to suppress target-target couplings \cite{farouk22}.}

\begin{acknowledgements}
The authors thank John Jeffers and Nick Spong for useful discussions and comments on the manuscript and M Squared Lasers for loan of equipment. This work was supported by funding from the UK National Quantum Technology Programme through ESPRC (Grant No. EP/N003527/1), the QuantIC Imaging Hub (Grant No. EP/T00097X/1), the University of Strathclyde {\color{black}{and}} QinetiQ. The data presented in the paper are available here \cite{mcdonnell21data}.
\end{acknowledgements}

%

\newpage{}

\pagebreak
\onecolumngrid
\vspace{\columnsep}
\begin{center}
\textbf{\large Supplemental Material for Demonstration of a Quantum Gate using Electromagnetically Induced Transparency}\\
\end{center}
\vspace{\columnsep}
\twocolumngrid


\section{Multi-level EIT model}
\subsection{EIT gate via $6P_{3/2}$}
\begin{figure}
    \centering
    \includegraphics{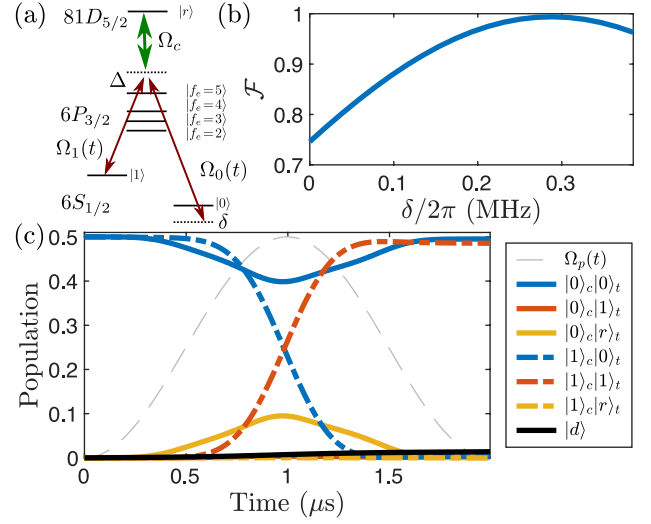}
    \caption{EIT Protocol via 6$P_{3/2}$ (a) Target qubit level scheme showing excitation to $\ket{81D_{5/2}, m_j=5/2}$ via $6P_{3/2}$ hyperfine states. (b) Optimisation of Raman transfer fidelity for $\ket{0}_t\rightarrow\ket{1}_t$ for $\Omega_c=0$ showing temporal AC shifts can be offset using a static detuning. (c) Bell state preparation for peak Raman power $P_\mr{p}=110~$nW focused to $w_0=3~\mu$m, $P_c=170$~mW focused to $w_0=18~\mu$m, $\Delta/2\pi=1.03$~GHz yielding $\mathcal{F}=0.98$.}
    \label{fig:BellFidelity6p32}
\end{figure}

To model the EIT on the target qubit for the realistic case of excitation via the $6P_{3/2}$ state with multiple hyperfine states we extend the simple four level model presented above as follows. For qubits encoded in the hyperfine clock states $\ket{4,0}$ and $\ket{3,0}$ the $\sigma^+$-polarized components of the Raman beams couple to states $\ket{f_e,1}$ in the excited state for $f_e=2-5$ as shown in Fig.~\ref{fig:BellFidelity6p32}(a). The Hamiltonian for the target qubit is given by
\begin{equation}
\begin{split}
\mathcal{H}_t/\hbar=&-\delta\ket{0}\!\bra{0}-(\Delta+\Delta_c)\ket{r}\!\bra{r}-\displaystyle\sum_{f_e}\Delta_{f_e}\ket{f_e}\!\bra{f_e}\\&+\displaystyle\sum_{f_e}\frac{1}{2}\left[\Omega_0^{f_e}(t)\ket{0}\!\bra{f_e}+\Omega_1^{f_e}(t)\ket{1}\!\bra{f_e}\right.\\
&\quad\left.+\Omega_c^{f_e}(t)\ket{r}\!\bra{f_e}+\mathrm{h.c.}\right],
\end{split}
\end{equation}
where $\delta$ is the two-photon detuning from $\ket{1}\rightarrow\ket{0}$, $\Delta$ is the centre of mass detuning from the intermediate state, $\Delta_{f_e}=\Delta-E_\mathrm{hfs}(f_e)$ is the detuning of the individual hyperfine states, $\Delta_c$ is the coupling laser detuning from the centre of mass detuning to $\ket{r}$, and $\Omega_{i}^{f_e}$ is the Rabi frequency for coupling from $\ket{i}\rightarrow\ket{f_e,m_{f_e}=1}$. 

Evolution of the system is calculated using a Lindblad master equation, including spontaneous emission from the excited states and Rydberg state at rates $\Gamma_e/2\pi=5.2$~MHz and $\Gamma_r/2\pi=1$~kHz respectively \cite{beterov09}. To account for the effect of leakage from the computational basis states we use the branching ratios from the excited state to calculate decay from $\ket{f_e}\rightarrow\ket{0,1}$, and assign the remaining decay amplitude and all decay from the Rydberg state to accumulate into an additional level $\ket{d}$ outside of the computational states.

\begin{table}[b!]
\caption{Summary of experimental Rabi frequencies corresponding to $\tau=2~\mu$s for $\Delta/2\pi=1.34$~GHz.\label{tab:Rabi}}
\begin{ruledtabular}
\begin{tabular}{l|cccc}
$\ket{f_e}$ & $\ket{2,1}$& $\ket{3,1}$& $\ket{4,1}$& $\ket{5,1}$\\
\hline
$\Omega_0^{f_e}/2\pi$~(MHz) & 26.1 & 42.3 & 26.6 & - \\
$\Omega_1^{f_e}/2\pi$~(MHz) & - & 14.1 & 37.3 & 39.9 \\
$\Omega_c^{f_e}/2\pi$~(MHz) & 17.8 & 38.4 & 43.6 & 27.1 \\
$\Delta_{f_e}/2\pi$~(GHz) & 1.474 & 1.322 & 1.121 & 0.870
\end{tabular}
\end{ruledtabular}
\end{table}

The Rabi frequencies for the probe transitions are scaled by $\Omega_i(t)=\Omega_i^\mathrm{max}[1-\cos(2\pi t/\tau)]/2$, however, due to the dipole selection rules only $f_e=3,4$ contribute to the two-photon Raman transfer from $\ket{1}\rightarrow\ket{0}$, whilst couplings to $f_e=2,5$ lead to additional AC Stark shifts and leakage due to spontaneous emission. Similarly, the EIT dark state is only created via interference of the excitation pathway via $f_e=3,4$, and direct two-photon resonances exist for $\ket{1}\rightarrow\ket{r}$ via $f_e=5$ and from $\ket{0}\rightarrow\ket{r}$ via $f_e=2$. Additional resonances to Rydberg states with $m_j=1/2,3/2$ are suppressed by choice of polarisation and by the 7.5~G quantisation field which are detuned by 25.2 and 12.1 MHz respectively, and can be neglected in the analysis. The effective two-photon Rabi frequency for Raman transitions is equal to $\Omega_R = \sum_{f_e}\Omega_1^{f_e}\Omega_0^{f_e}/2\Delta_{f_e}$, which is related to the pulse duration as $\tau=8\pi/(3\Omega_R)$. Table~\ref{tab:Rabi} summarizes peak Rabi frequencies for the experimental parameters chosen for $\tau=2~\mu$s with a peak Raman power of 110~nW focused to $w_0=3~\mu$m giving $\Omega_R/2\pi=0.67$~MHz and a coupling laser power 170~mW focused to $w_0=18~\mu$m.

Differential shifts on the Raman transition are equal to $\delta_\mathrm{AC}=\sum_{f_e}[(\Omega_1^{f_e})^2-(\Omega_0^{f_e})^2]/4\Delta_{f_e}+(\Omega_1^{f_e})^2/4(\Delta_{f_e}+\omega_q)-(\Omega_0^{f_e})^2/4(\Delta_{f_e}-\omega_q)$ where $\omega_q/2\pi=9.2$~GHz is the {\color{black}{qubit hyperfine-splitting}}, which can be compensated through tuning $\delta$. Whilst these AC shifts are time-dependent, it is possible to compensate using a fixed Raman detuning, as shown in Fig.~\ref{fig:BellFidelity6p32}(b) where we model fidelity of transfer from $\ket{1}_t\rightarrow\ket{0}_t$ as a function of $\delta$ for the case $\Omega_c=0$. This yields a peak transfer error of $1-\mathcal{F}=6\times10^{-3}$ at $\delta/2\pi = 0.28$~MHz.

To evaluate the intrinsic Bell state preparation fidelity we model the case of an initial state $\ket{\psi}=(\ket{0}_c+\ket{1}_c)/\sqrt{2}\otimes\ket{0}_t$ and calculate the fidelity of creating the $\ket{\Phi^+}$ Bell state after application of the EIT gate protocol. The interaction is included as $\mathcal{H'}=\hbar V(R)\ket{rr}_t\bra{rr}$ where $V(R)/2\pi = 34.9$~MHz. The results of the simulation are shown in Fig.~\ref{fig:BellFidelity6p32}(c), resulting in a Bell state fidelity $\mathcal{F}=0.98$, with the dominant error due to leakage out of the computational basis due to the finite lifetime of the control and target Rydberg states resulting in $\rho_d=0.013$. A second significant error channel introduced in the experiment when using a common coupling laser to illuminate both control and target atoms is scattering of light from the control atom due to the finite intermediate state detuning. Repeating simulations with the coupling laser applied to the control qubit during the pulse sequence we find this leads to an increase in population of $\ket{d}$ to $\rho_d=0.047$ and a reduction in fidelity to $\mathcal{F}=0.95$. This limitation can be overcome in future using either single-site addressing of the coupling laser onto the target qubits, or for parallel gate implementation using a dual-species approach to ensure the control atoms do not scatter light from the Rydberg laser acting on target qubits \cite{beterov15}.

\subsection{EIT gate via the $7P_{1/2}$ transition}

\begin{figure}[t!]
    \centering
    \includegraphics{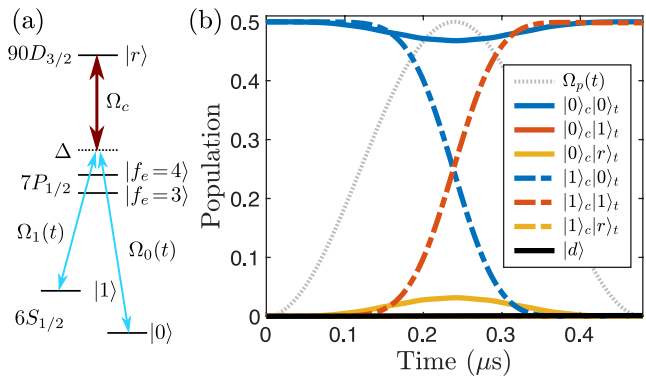}
    \caption{(a) Target qubit level scheme showing excitation to $\ket{90D_{3/2}, m_j=3/2}$ via $7P_{1/2}$. (b) Bell state preparation for peak Raman power $P_\mr{p}=200~\mu$W, $P_c=50$~mW focused to $w_0=3~\mu$m, $\Delta/2\pi=5$~GHz yielding $\mathcal{F}>0.998$.}
    \label{fig:BellFidelity7p12}
\end{figure}

Improvements to the performance of the EIT gate protocol can be obtained by changing from a $P_{3/2}$ to $P_{1/2}$ intermediate state to suppress the $f_e=2,5$ hyperfine states that do not contribute to the EIT and Raman pathways but cause leakage from off-resonant scattering from both Raman and coupling lasers. Additionally switching to the inverted excitation scheme using excitation via the $7P_{1/2}$ state as shown in Fig.~\ref{fig:BellFidelity7p12}(a) provides two further advantages. Firstly, a $\times1/5$ reduction in natural linewidth ($\Gamma_e/2\pi=1.03$~MHz) and secondly this means the transition from $\ket{e}\rightarrow\ket{r}$ changes from a visible to near-infrared wavelength at 1039~nm where high power fiber lasers are available. This, combined with the stronger Rydberg transition matrix elements enables $\Omega_c$ to be maximized, permitting fast gate operations and increased intermediate state detuning, providing scope to increase the number of target qubits. 

Using the model presented above, we simulate the EIT gate performance for excitation to the $\ket{90D_{3/2}, m_j=3/2}$ Rydberg state from $7P_{1/2}$ using a probe power of $200~\mu$W shared equally between the two Raman components and a coupling power of 50~mW, focused onto {\color{black}{each qubit}} with a $1/e^2$-radius of 3~$\mu$m. For an intermediate state detuning of $\Delta/2\pi=5$~GHz this corresponds to an adiabatic pulse duration of 500~ns as shown in Fig.~\ref{fig:BellFidelity7p12}(b), with a Bell state fidelity of $\mathcal{F}>0.998$. 

{\color{black}{
\section{Multi-qubit gate scaling}

As described in the introduction, the advantage of the EIT gate protocol is its scalability to multi-qubit gate operations. Using the parameters for the optimised two-qubit gate protocol shown in Fig.~\ref{fig:BellFidelity7p12}, we consider scaling of gate fidelity as a function of qubit number $k$ by calculating the fidelity of mapping initial state $(\ket{0}_c+\ket{1}_c)/\sqrt{2}\otimes\ket{0}_t^{\otimes k}$ to the GHZ state $\ket{0}_c\ket{0}_t^{\otimes k}-(-1)^k\ket{1}_c\ket{1}_t^{\otimes k})/\sqrt{2}$ following application of the CNOT$^k$ gate protocol. This factor of $(-1)^k$ arises from the multiplication of the $\pi$-phase associated with ground-state transfer on each target qubit.

\begin{figure}[t!]
    \centering
    \includegraphics{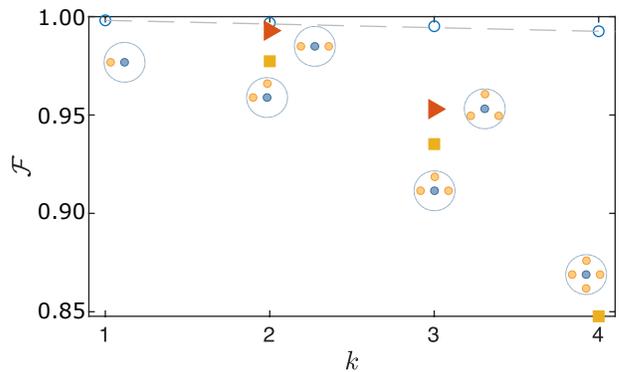}
    \caption{\color{black}CNOT$^k$ fidelity vs $k$ for preparing GHZ states using the EIT gate with optimal parameters identified in Fig.~\ref{fig:BellFidelity7p12}. Filled points are modelled with real target-target interactions $V(R)\propto1/R^6$, whilst open circles have target-target couplings suppressed. For each filled point the relative geometry of the control and target qubits are illustrated. Dashed line indicates expected scaling in the absence of target-target interactions with $\mathcal{F}_k=(\mathcal{F}_1)^k$.}
    \label{fig:multiQ}
\end{figure}
\color{black}
We model the process up to $k=4$ as required for protocols based on surface-code error correction \cite{auger17}, using optimal pulse parameters for excitation via 7P$_{1/2}$ used above. Results are shown in Fig.~\ref{fig:multiQ} for the case of both realistic target-target interactions (filled points) and neglecting target-target interactions (open points). For each case we consider different geometric configurations (as shown) which maintain a spacing of $R=4~\mu$m between control and target qubits, and scale the target-target couplings appropriately ($\propto 1/R^6$). These results show that for $k=2$ a line geometry with a central control qubit out performs a right-angled configuration as expected, with a big drop in fidelity for $k>2$ when the target-target distance is reduced below 8~$\mu$m. In the absence of target-target interactions the gate scales as $\mathcal{F}_k=(\mathcal{F}_1)^k$, with CNOT$^4$ yielding a fidelity $\mathcal{F}>0.899$ equivalent to performing four sequential CNOT gates with $\mathcal{F}_1>0.998$. Using a single Rydberg level on both control and target however, the residual target-target couplings limit performance by compromising the EIT dark state. Robust scaling towards multi-qubit gates thus requires judicious choice of Rydberg states to maximise control-target couplings whilst suppressing target-target interactions, possible using either different Rydberg levels of the same species or using Rydberg states of different atomic species \cite{beterov15,farouk22}, but the non-interacting results show the feasibility of scaling of this approach.
}}
\section{Experiment Details}

\subsection{Optical Pumping Procedure}

We use the Cs $D_1$ line to optically pump atoms into qubit state $\ket{1}$ using $\pi$-polarized light on the transition $\ket{6S_{1/2}, F = 4} \rightarrow \vert 6P_{1/2}, F^{'} = 4\rangle$ but find that the optical pumping efficiency is limited by polarisation purity to 95\% due to stress-induced birefringence on the viewport window. This finite preparation purity means there is a small fraction of the atomic population in $\ket{6S_{1/2}, F = 4, m_F\neq0}$ states which are outside of the computational basis. This introduces large errors when performing the EIT gate protocol due to weak excitation of these states to the Rydberg manifold interfering with the adiabatic evolution on the target site.

To suppress these errors, after optical pumping into $\ket{11}$ we apply a resonant microwave $\pi$-pulse to the atoms, followed by a blow-away pulse resonant with $\ket{6S_{1/2}, F = 4} \rightarrow \vert 6P_{3/2}, F^{'} = 5\rangle$ to remove any atoms remaining in $F=4$, and provides a high-purity preparation of $\ket{00}$ with a probability of 90(1) \%. Whilst this reduces the overall atom retention, this is a constant error which can be independently measured for rescaling to perform background correction.

\subsection{Microwave State Preparation}
Following optical pumping we perform state preparation using ground state microwave rotations to provide a robust method for implementing high-fidelity single qubit gates from $\ket{0}\rightarrow\ket{1}$ free from errors of spontaneous emission that are present using Raman transitions. {\color{black}{To start, atoms are optically pumped into $\ket{11}$, then transferred to $\ket{00}$ using a global microwave $\pi$-pulse $X(\pi)$.}} However due to the comparatively long wavelength ($\lambda\sim$1~cm) {\color{black}{the microwave field}} cannot be used to spatially address a single atom. To implement local microwave rotations, we follow the method of Xia \emph{et al}.~\cite{xia15} using the qubit laser $\Omega_a$ focused on the control atom to apply a differential AC Stark shift that detunes the atom from the microwave resonance without disturbing the target atom. For a resonant microwave rotation area $\theta = \Omega t$ on the target atom, the AC Stark shift $\Delta'$ is chosen to give a rotation $\theta^{'} = \Omega^{'}t = 4\pi$ on the control atom, where  $\Omega^{'}=\sqrt{\Omega^2 + \Delta'^2}$, leaving the qubit state of the control atom unchanged. Re-arranging gives 
\begin{equation}
    \frac{\abs{\Delta'}}{\Omega} = \sqrt{\frac{16 \pi^2}{\theta^2}-1},
\end{equation}
which for a microwave $\pi$-pulse {\color{black}{gives}} $\abs{\Delta'_\pi} = \sqrt{15}\Omega$. For the measured microwave Rabi frequency of $\Omega/2\pi =3.31(6)$~kHz, this corresponds to an applied shift of $\abs{\Delta_\pi}/2\pi = 12.8$~kHz, which is calibrated independently using Ramsey spectroscopy.

Using this technique, we prepare state $\ket{01}$ using
\begin{center}
\begin{tikzcd}
\lstick{$\ket{0}_c$} & \qw &  \gate{X(4\pi)}  & \qw    & \rstick{$\ket{0}_c$} \qw \\
\lstick{$\ket{0}_t$} & \qw &  \gate{X(\pi)}  & \qw    & \rstick{$\ket{1}_t$} \qw
\end{tikzcd}
\end{center}
where $X(4\pi)$ reflects the effect of a local AC shift on the target site, whilst $\ket{10}$ is achieved using
\begin{center}
\begin{tikzcd}
\lstick{$\ket{0}_c$} &\qw &  \gate{X(\pi)}& \qw &  \gate{X(4\pi)}  & \qw    & \rstick{$\ket{1}_c$} \qw \\
\lstick{$\ket{0}_t$} & \qw &  \gate{X(\pi)}& \qw &  \gate{X(\pi)}  & \qw    & \rstick{$\ket{0}_t$} \qw `
\end{tikzcd}
\end{center}
In a similar manner, to rotate the output states back to $\ket{00}$ for measurement we use the reverse sequence. 

\subsection{State Readout}
For neutral atom qubits, the dominant error in state detection and readout is the inability to distinguish atom loss from measurement of an atom in state $\ket{1}$. Following Refs.~ \cite{gaetan10,levine19} we introduce the state $\{\ket{x}\}$ corresponding to the case of qubits being lost during the gate sequence, or those outside the computational basis either due to imperfect state preparation or leakage from the excitation pulses. We can now relate these states to measurements in the two-atom basis which can be performed either using a state selective blow-away beam to expel atoms in $\ket{1}$ from the trap ($A$), or without ($B$). For both cases we obtain four possible outcomes $\{\bullet\bullet,\bullet\circ,\circ\bullet,\circ\circ\}$ where $\bullet$ corresponds to an atom present and $\circ$ an atom loss. 

When using blow away, the measurements correspond to the operators
\begin{subequations}
\begin{align}
A_{\bullet\bullet} & = \rho_{00},\\
A_{\bullet\circ} & = \rho_{01}+\rho_{0x},\\
A_{\circ\bullet} & = \rho_{10}+\rho_{x0},\\
A_{\circ\circ} & = \rho_{11}+\rho_{1x}+\rho_{x1}+\rho_{xx},
\end{align}
\end{subequations}
whilst for measurements taken without blow-away we obtain
\begin{subequations}
\begin{align}
B_{\bullet\bullet} & = \rho_{00}+\rho_{01}+\rho_{10}+\rho_{11},\\
B_{\bullet\circ} & = \rho_{0x}+\rho_{1x},\\
B_{\circ\bullet} & =\rho_{x0}+\rho_{x1},\\
B_{\circ\circ} & = \rho_{xx},
\end{align}
\end{subequations}
where $\rho_{ij}=\ket{ij}\!\bra{ij}$ are the projection operators.

As can be seen from these equations, $A_{\bullet\bullet}$ is insensitive to single atom losses and thus provides a robust output measurement state, whilst $B_{\bullet\bullet}$ provides a measurement of the total population within the computational basis. Following measurement of both populations, the loss-corrected probability is extracted using $\rho_{00}^\mathrm{cor}=A_{\bullet\bullet}/B_{\bullet\bullet}$.

{\color{black}{\subsection{Cross-Talk Error}
The gate procedure relies on independent addressing of control and target qubits. Experimentally this is achieved by focusing lasers down onto a single qubit using $1/e^2$ waists of $3~\mu$m (recall the separation of the qubits is $R=6~\mu$m), which results in a factor of $\exp(-8)$ intensity leakage on the qubits not being addressed, and a theoretical cross talk error below $1\times10^{-3}$. Experimentally we have measured zero probability of transferring either the control qubit to $\ket{0}$ when applying a Raman $\pi$-pulse on the target or the target qubit to the Rydberg state after a $\pi$-pulse on the control qubit after 200 measurements, bounding the error below $9\times10^{-3}$ limited by the number of measurements used to determine cross talk.}}

\subsection{CNOT Gate State Preparation and Measurement}
To characterize the CNOT gate performance we first prepare the qubits in each of the computational basis states using microwave rotations as described above, then following the gate sequence apply further microwave pulses to rotate each of the possible computational basis states back into $\ket{00}$. For each input and output combination, we repeat the experiment with ($A$) and without ($B$) blow away from which we extract $A_{\bullet\bullet}$ to obtain the raw (uncorrected) gate elements, and normalise by the corresponding $B_{\bullet\bullet}$ to obtain the loss-corrected output probabilities. {\color{black}{Error bars are calculated using $\delta P = \sqrt{P(1-P)/n}$ where $n$ is the number of times both atoms loaded. Measurements are repeated 200 times for each element, with an average two-atom load probability $\sim25\%$.}}

\begin{figure}[t!]
    \centering
    \includegraphics{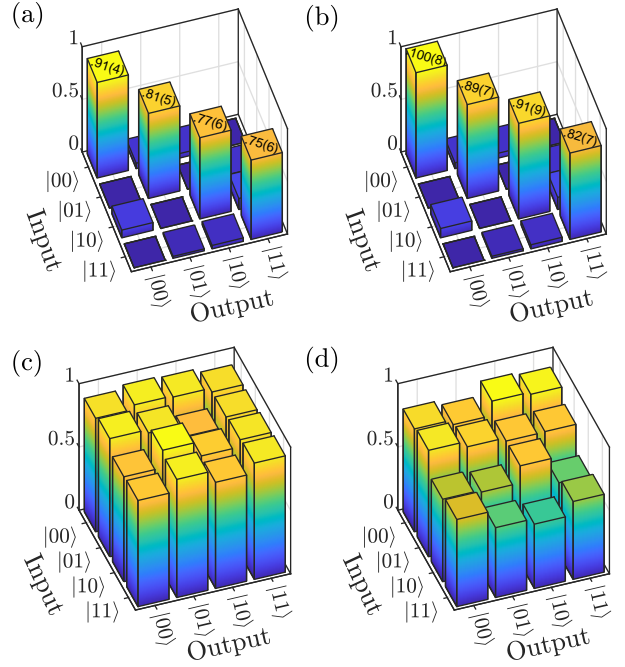}
    \caption{State preparation and normalisation data. (a) Raw and (b) loss-corrected data for the state preparation matrix yielding $\mathcal{F}_\mathrm{prep}=0.81(2)$ and $\mathcal{F}_\mathrm{prep}^\mathrm{cor} = 0.91(4)$ respectively, with normalisation data measured without blow away shown in (c). (d) Normalisation data for CNOT gate data presented in Fig. 4(b) of the main paper. This shows that for the EIT gate the average two-atom survival when control qubit is in state $\ket{0}_c$ is $\langle B_{\bullet\bullet}\rangle=0.82(2)$ whilst for the control in $\ket{1}_c$ it reduces by 25\% to $\langle B_{\bullet\bullet}\rangle=0.62(2)$ due to losses from exciting to the Rydberg state.}
    \label{fig:normalize}
\end{figure}

Results for the uncorrected ($A_{\bullet\bullet}$) and corrected ($A_{\bullet\bullet}/B_{\bullet\bullet}$) state preparation matrices (measured without gate pulses) are presented in Fig.~\ref{fig:normalize}(a)-(b). State preparation fidelities are calculated using $\mathcal{F}_\mathrm{prep}=1/4\mathrm{Tr}(U^T_\mathrm{meas}*U_\mathrm{ideal})$, with the raw data (a) giving 0.81(2) and corrected date (b) giving $\mathcal{F}_\mathrm{prep}^\mathrm{cor}=0.91(4)$. Normalisation data measured without a blow-away pulse ($B_{\bullet\bullet}$) are shown in (c), corresponding to an average two-atom survival probability of 0.89(1). Also shown in Fig.~\ref{fig:normalize}(d) is the normalisation data used to correct the CNOT gate matrix in Fig.~4 of the main paper, {\color{black}{which clearly shows around 25\% reduction in two-atom survival when the control atom is in state $\ket{1}_c$. This additional loss is a result of the control qubit not returning from the Rydberg state, which is verified by comparing the probability of the control qubit surviving when the target atom is not present. With no gate pulses, we measure a single atom survival probability of $0.92(3)$, whilst following the $\pi$-gap-$\pi$ Rydberg pulses on the control qubit with a gap time of $\tau=2~\mu$s we observe a reduced survival of 0.70(4) which is consistent with the 25\% loss observed in the experiments. An intrinsic source of loss from the Rydberg level arises from spontaneous decay or off-resonant scattering from the coupling laser, which was estimated in the theory section above to account for approximately 5\% loss. The remaining loss is attributed to laser phase noise which causes dephasing, and causing imperfect return of population back to state $\ket{1}_c$ after the second $\pi$-pulse \cite{levine18}. This is consistent with the observe a $1/e$ damping time of 1.5~$\mu$s when driving the control qubit to the Rydberg state.}} 

\section{Bell State Parity Analysis}
We characterize the experimentally prepared state $\rho$ by evaluating its fidelity with respect to the target Bell state $F = \bra{\Phi^+}\rho\ket{\Phi^+}=(\rho_{00}+\rho_{11})/2+\rho_{00,11}/2$. The fidelity is the sum of two terms, the first of which are the Bell state populations, equal to the probability of observing $\ket{00}$ or $\ket{11}$. The second term is the coherence between $\ket{00}$ and $\ket{11}$ equal to $\rho_{00,11}=\vert c\vert e^{i\phi_c}$. To extract the coherence, we allow the state to accumulate a global phase shift $Z(\phi)$ followed by applying a global $X(\pi/2)$ rotation on both qubits, and record the parity $\Pi=\rho_{00}+\rho_{11} - \rho_{01} - \rho_{10}$ as a function of $\phi$.

In order to understand the impact of losses on the parity measurements, we consider each qubit to have states $\ket{0},\ket{1}$ and $\ket{x}$ where $x$ denotes an atom that is outside of the computational basis. For an arbitrary initial state $\rho$, following application of the global $Z(\phi)$ and $X(\pi/2)$ rotations we evaluate the output of the operator 
\begin{equation}
\hat{\tilde{\Pi}}=\hat{A}_{\bullet\bullet}+\hat{A}_{\circ\circ}-\hat{A}_{\circ\bullet}-\hat{A}_{\bullet\circ},
\end{equation}
using the definitions of $\hat{A}_{ij}$ from above, which in the absence of loss is exactly equivalent to the parity $\Pi$. This results in an experimentally measured parity of 
\begin{equation}
\tilde{\Pi}(\phi)=2\Re(d)-2\vert c\vert \cos(2\phi+\phi_c)+\rho_{xx},
\end{equation}
corresponding to a parity oscillation at frequency $2\phi$ with amplitude $\vert c\vert$ with an offset dependent on the two-atom loss probability $\rho_{xx}$ and the coherence associated with the alternative Bell states which are included as $\rho_{10,01}=\vert d\vert e^{i\phi_d}$. These results show the measurement of parity oscillations using the global pulse sequence is robust to errors from atom loss, and the observed amplitude $\vert c\vert$ can be re-normalized against the averaged two-qubit survival without blow away $\langle B_{\bullet\bullet}\rangle_\phi$.

\end{document}